\documentclass[sigconf, nonacm]{acmart}

\AtBeginDocument{%
  }

\usepackage{subcaption}
\usepackage{graphicx}

\makeatletter
\def\blfootnote{\xdef\@thefnmark{}\@footnotetext}
\makeatother

\begin{document}

\title{Membership Inference over Diffusion-models-based Synthetic Tabular Data}


\author{Peini Cheng}
\affiliation{%
  \institution{University of Alberta}
  \city{Edmonton}
  \state{Alberta}
  \country{Canada}}

\author{Amir Bahmani}
\affiliation{%
  \institution{University of Alberta}
  \city{Edmonton}
  \state{Alberta}
  \country{Canada}}

\begin{abstract}
    This study investigates the privacy risks associated with diffusion-based synthetic tabular data generation methods, focusing on their susceptibility to Membership Inference Attacks (MIAs). We examine two recent models, TabDDPM and TabSyn, by developing query-based MIAs based on the step-wise error comparison method. Our findings reveal that TabDDPM is more vulnerable to these attacks. TabSyn exhibits resilience against our attack models. Our work underscores the importance of evaluating the privacy implications of diffusion models and encourages further research into robust privacy-preserving mechanisms for synthetic data generation.
\end{abstract}
\maketitle

\blfootnote{This work is a course project under the supervision of Dr. Bailey Kacsmar.}

\section{Introduction}
The availability of large training datasets and recent advances in machine learning have demonstrated great potential to benefit communities across various sectors, such as healthcare and education \cite{holmes2019artificial, rajkomar2019machine}. However, this potential is significantly limited by privacy concerns associated with sharing sensitive data with researchers and organizations. It has been shown that common methods of de-identifying personal records are prone to re-identification, especially when different datasets are combined \cite{Sweeney1997, Narayanan2008, ElEmam2011}. Despite efforts to improve de-identification techniques, the risk of re-identification persists, particularly in datasets with rich feature sets that provide numerous points for potential cross-linking \cite{ElEmam2011}. In addition to this, if the data can be associated with any natural person's identity, then the sharing of the large training datasets is often limited by regulations like the European General Data Protection Regulation (GDPR) as a protection of human rights. As outlined in Article 7 of the GDPR, explicit consent from data subjects is required for data processing and they have the right to withdraw it at any time, which introduces great difficulties for researchers. These challenges have encouraged a broader exploration of data generation and sharing techniques that balance utility with privacy compliance, with an increasing emphasis on solutions that align with GDPR Recital 26, which excludes anonymous data from the scope of data protection requirements \cite{Jordon2019}.

Synthetic data generation addresses this challenge by providing a way to share data that maintains the statistical properties of the original dataset without exposing sensitive individual information. This approach works because we don't need the actual data; it is sufficient for the synthetic data to resemble the real data, which is adequate for training machine learning models. Moreover, synthetic data's applications extend beyond privacy, aiding in scenarios like rare event modeling, data augmentation, and handling imbalanced datasets, significantly enhancing machine learning robustness \cite{zhao2022ctabganenhancingtabulardata}. By not posting the real data, the processing of the data would potentially be free from being governed by GDPR, as it is outlined in Recital 26 of the GDPR that the principles of data protection is not applicable to the processing of anonymous data. Generative models trained on original data have proven effective in generating high quality synthetic data that preserves the statistical properties of the original dataset while enhancing privacy. Furthermore, efforts have been made to integrate these models with differential privacy to provide formal privacy guarantees \cite{Jordon2019}. However, the implementation of differential privacy in generative models often involves trade-offs, such as reduced data utility or increased computational complexity, underscoring the need for innovative solutions that maintain this delicate balance \cite{pang2024clavaddpmmultirelationaldatasynthesis}.

Recent developments in diffusion models have attracted great attention due to their effectiveness across various data types. These models have also shown remarkable capability in synthesizing high quality tabular data that preserves the complex relationships present in real world datasets. \textbf{TabDDPM} provides a simple design of Denoising Diffusion Probabilistic Models for tabular problems that can be applied to various tasks and works effectively with mixed data types. \textbf{TabSyn} leverages diffusion models but takes a different approach by employing a Variational Autoencoder (VAE) to encode tabular data into a latent space, then utilizing a diffusion generative model to learn and generate data from the latent distribution \cite{ zhang2024mixedtypetabulardatasynthesis, kotelnikov2022tabddpmmodellingtabulardata}.

Although it seems that sharing synthetic data can bypass compliance with regulations, there is still the risk of these models inadvertently leaking real data from the training set. Distance to Closest Record (DCR) is a commonly used privacy evaluation metric that quantifies the similarity between the synthetic data and the real data \cite{kotelnikov2022tabddpmmodellingtabulardata, pang2024clavaddpmmultirelationaldatasynthesis}. This evaluation method can be directly applied to any generative model since it only requires real and synthetic datasets and does not require access to the model. However, it does not account for attacks targeting the model such as the Membership Inference Attacks (MIAs), which have been proposed as an effective tool to measure the privacy risk of data leakage \cite{murakonda2020mlprivacymeteraiding}. The goal of the adversary performing MIAs is to determine whether a specific individual's data was used in the training dataset of a machine learning model \cite{shokri2017membershipinferenceattacksmachine}. The success of an MIA poses a significant privacy risk, since it provides the advantage for the attacker to perform other attacks, including Attribute Inference Attacks (AIAs). With the auxiliary knowledge of that individual, AIAs can expose sensitive information from the training set.

\section{Problem Statement}
The problem we aim to address in our study is finding a quantitative evaluation of the data breach risk generated by diffusion-based synthetic tabular data methods. Specifically, we focus on the resistance of our target models to MIAs. Such evaluations are particularly critical because even minor vulnerabilities can lead to significant privacy breaches, which, in turn, may undermine the trust in synthetic data adoption and regulatory compliance. Despite, recent advances in diffusion models, their privacy resilience has remained unexplored. Diffusion models, due to their iterative noise-based data transformation processes, inherently introduce complexity in privacy evaluation, particularly when assessing the trade-off between reconstruction fidelity and data privacy risks. We examine the privacy of two recent diffusion model-based tabular approaches TabDDPM, TabSyn by developing query-based membership inference attacks, and evaluating the success of our attacks by their ability to determine whether a data point was used for training the generative model.

\section{Related work}
\textbf{Synthetic tabular data generation} is increasingly viewed as a promising approach to address challenges in privacy-compliant data sharing. The machine learning research community is actively developing models tailored for generating tabular data. Several studies have created GAN-based approaches, \citeauthor{Park_2018} introduced TableGAN which consists of three neural networks - a generator neural network that produces data with similar distribution, a discriminator neural network that classifies data as synthetic or real , and a classifier neural network that prevent the generation of semantically incorrect data. It is capable of generating tables that contain categorical, discrete, and continuous values, however, it cannot deal with columns with mixed data type. \citeauthor{zhao2021ctabganeffectivetabledata} introduced CTAB-GAN that overcomes this challenge with additional Mixedtype Encoder and outperformed the previous approach. Later, they introduced CTAB-GAN+ \cite{zhao2022ctabganenhancingtabulardata} with even better utility and differential privacy guarantees. While GAN-based models like CTAB-GAN+ demonstrate advanced capabilities in data synthesis, they face challenges in addressing mode collapse and ensuring diversity in the generated samples, which remain critical for real-world applications \cite{zhao2021ctabganeffectivetabledata}.

\textbf{Membership inference attack} was first introduced by \citeauthor{shokri2017membershipinferenceattacksmachine}. The attack begins by training a set of shadow models that imitate the behavior of the target model. These shadow models are trained on a dataset fully known to the attacker. By using both the true labels from this dataset and the predictions made by the shadow models, the attacker then trains an attack model that learns to classify whether the prediction belongs to a member of training set. \citeauthor{hayes2018loganmembershipinferenceattacks} then proposes GAN-Leaks against generative adversarial networks (GANs). Assuming the attacker have access to the generator, they collect samples from the generator and approximate the probability of the specific sample being generated. Higher probability implies a higher likelihood of membership. However, GAN-Leaks of white-box settings is computationally infeasible for diffusion models, and the black-box version does not achieve strong performance in attacking diffusion models, as reported by \citeauthor{duan2023diffusionmodelsvulnerablemembership}. Therefore, they designed a MIA approach specifically for diffusion models, which is explained in Section \ref{secmi}.

\section{Background}
\subsection{Diffusion-models-based synthetic tabular data}
\begin{figure}[ht]
\centering
\includegraphics[width=1\linewidth]{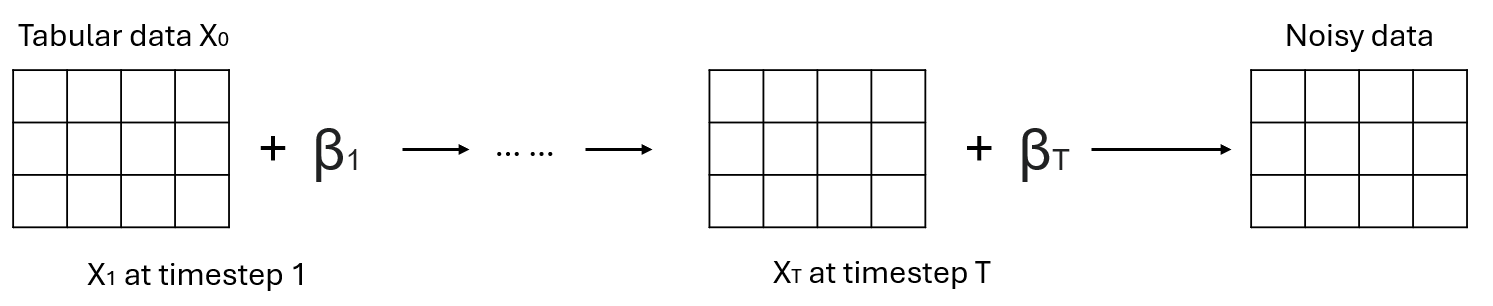}
\caption{\label{fig:forward} Forward process of the diffusion model. Scheduled noise \begin{math}\beta_t\end{math} is gradually added until the data is close to pure noise.}
\end{figure}

Diffusion models are latent variable models that define the forward diffusion process and the reverse process as a Markov chain\cite{ho2020denoisingdiffusionprobabilisticmodels}. The forward process gradually adds scheduled Gaussian noise \begin{math}\beta_t\end{math} at each timestep t to the input tabular data (see Figure \ref{fig:forward}). This transition at timestep t is denoted as \begin{math} q(x_t|x_{t-1})\end{math}:
\begin{equation}
q(x_t|x_{t-1}) = \mathcal{N}(x_t;\sqrt{1-\beta_t}x_{t-1},\beta_t\boldsymbol{I}) 
\end{equation}

\begin{figure}[ht]
\centering
\includegraphics[width=0.8\linewidth]{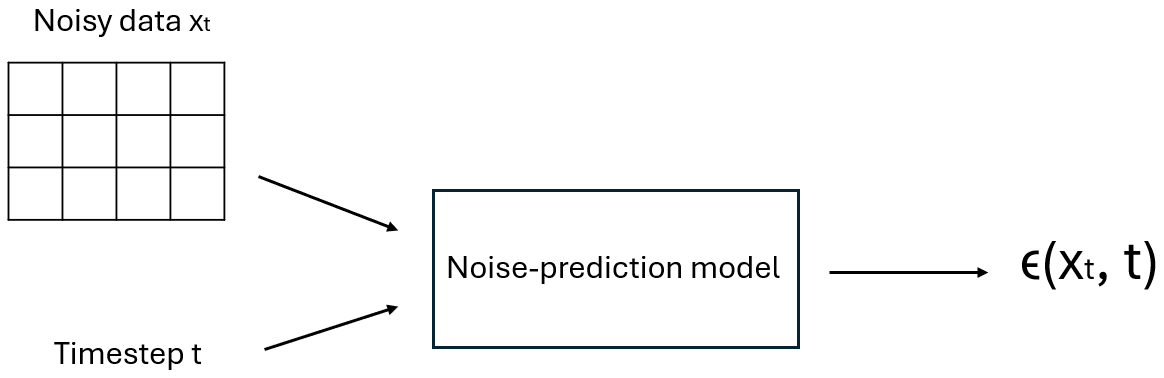}
\caption{\label{fig:epsilon} Neural network \begin{math}\epsilon_\theta(x_t,t)\end{math} in the diffusion model that takes the timestep t and the noisy data at timestep t as input and estimates the noise added at timestep t. Also known as the denoising model.}
\end{figure}

Synthetic data are generated by the reverse process denoted by \begin{math} p_\theta(x_t|x_{t-1})\end{math}:
\begin{equation}
p_\theta(x_{t-1|x_t}) = \mathcal{N}(x_{t-1};\mu_\theta(x_t,t),\Sigma_\theta(x_t,t))
\end{equation}

where 
\begin{equation}
\mu_\theta(x_t,t) = \frac{1}{\sqrt{\alpha_t}}(x_t - \frac{\beta_t}{\sqrt{1 - \bar{\alpha}_t}}\epsilon_\theta(x_t,t))
\end{equation}

where \begin{math}\alpha_t = 1-\beta_t\end{math},  \begin{math}\bar{\alpha}_t = \prod^t_{i=1} \alpha_i\end{math} and \begin{math}\epsilon_\theta(x_t,t)\end{math} is a neural network with parameters \begin{math}\theta\end{math} trained to estimate the Gaussian noise at timestep t (See Figure \ref{fig:epsilon}). Training is performed by optimizing the variational bound which can be simplified to:
\begin{equation}
\label{eq:mse}
\ell_{x_0, t} = \mathbb{E}_{x_0, t} \left\| \epsilon - \epsilon_\theta(x_t, t) \right\|_2^2
\end{equation}

\subsection{Step-wise Error Comparing Membership Inference Attack} \label{secmi}
Assume we do not know the training dataset for the pre-trained model, but we do have access to a public dataset that contains the same type of data (e.g., same columns) and possibly a portion of the training dataset. The training data of the target model are defined as the member, while other data from the same public dataset are the non-member. Our goal as the attacker is to correctly infer the membership as much as possible.

\citeauthor{duan2023diffusionmodelsvulnerablemembership} made the assumption that the sum of mean-squared errors (Equation \ref{eq:mse}) of a member data is more likely to be smaller than a non-member data, if the model has overfitting issue. However, the calculation of that equation is not feasible. Therefore, they proposed t-error as an estimation:
\begin{equation}
\label{eq:terror}
\tilde{\ell}_{t, x_0} = \left\| \psi_\theta \left( \phi_\theta (\tilde{x}_t, t), t \right) - \tilde{x}_t \right\|^2
\end{equation}

where
\begin{equation}
\label{eq:deterministic}
\begin{split}
x_{t+1} &= \phi_\theta(x_t, t)\\
&= \sqrt{\bar{\alpha}_{t+1}} f_\theta(x_t, t) + \sqrt{1 - \bar{\alpha}_{t+1}} \epsilon_\theta(x_t, t) \\
x_{t-1} &= \psi_\theta(x_t, t) \\
&= \sqrt{\bar{\alpha}_{t-1}} f_\theta(x_t, t) + \sqrt{1 - \bar{\alpha}_{t-1}} \epsilon_\theta(x_t, t)
\end{split}
\end{equation}
where
\begin{equation}
\label{eq:f}
f_\theta(x_t, t) = \frac{x_t - \sqrt{1 - \bar{\alpha}_t} \epsilon_\theta(x_t, t)}{\sqrt{\bar{\alpha}_t}}.
\end{equation}

They demonstrated their assumption by showing that the t-errors of non-member is higher than member. It is worth noting that this difference is only significant at small timestep t, possibly because both data would be very noisy at a large timestep t that no meaningful difference between member and non-member data can be observed.

They also proposed two ways of attack: \begin{math}SecMI_{stat}\end{math} and \begin{math}SecMI_{NNs}\end{math}

For \begin{math}SecMI_{stat}\end{math}, they calculate a threshold to distinguish the member and non-member purely based on their t-errors. In this attack, a piece of data with a t-error lower than the threshold will be labeled as a member, while those with higher t-error will be labeled as non-member. For \begin{math}SecMI_{NNs}\end{math}, an attack model is implemented to predict membership scores with their t-errors. It was trained on selected samples' t-errors and their labels (whether the t-error belongs to a member). Details on these attack approaches are described in Section \ref{attack}. 

\section{Methodology}
We follow the assumption that the attacker has the access to the denoising model, as if the model was published to a public platform.
The dataset in Section \ref{dataset} will be split into two disjoint datasets: one being our member dataset, the other being the non-member dataset. The target model will then be trained solely on the member dataset which is known to us.

\subsection{Dataset} \label{dataset}
To test the effectiveness of our method against various types of tabular data, two public datasets will be used for training and evaluating target models \{ \textbf{Default}\footnote{\url{https://archive.ics.uci.edu/dataset/350/default+of+credit+card+clients}}, 
\textbf{Shoppers}\footnote{\url{https://archive.ics.uci.edu/dataset/468/online+shoppers+purchasing+intention+dataset}}\}. The information of these two dataset is displayed in Table \ref{tab:dataset}.
These two datasets have already been used to test their privacy protection ability by calculating Distance to Closet Record \cite{zhang2024mixedtypetabulardatasynthesis}. \citeauthor{zhang2024mixedtypetabulardatasynthesis} claimed that both TabDDPM and TabSyn do not suffer from privacy issues, with TabSyn demonstrating better privacy protection than TabDDPM. Through our experiments with MIAs, we will verify whether their claim is accurate.

\begin{table}
    \centering
    \begin{tabular}{|c|c|c|}
        \hline
         & Shoppers & Default \\
         \hline\hline
         \#Rows & 12,330 & 30,000\\
         \hline
         \#Num & 10 & 14\\
         \hline
         \#Cat & 18 & 11\\
         \hline
    \end{tabular}
    \caption{Statistics of datasets. \#Rows is the number of rows in the dataset, \#Num is the number of numerical columns, and \#Cat is the number of categorical columns.}
    \label{tab:dataset}
\end{table}

\subsection{Target model} \label{target}
SecMI was shown to be effective for attacking both Denoising Diffusion Probabilistic Models (DDPM)  and Latent Diffusion Models (LDM). Therefore, we adopt this approach to evaluate the privacy risks of two diffusion-based synthetic tabular data methods: TabDDPM\cite{kotelnikov2022tabddpmmodellingtabulardata}, a DDPM, and TabSyn\cite{zhang2024mixedtypetabulardatasynthesis}, an LDM. Note that TabSyn generates numerical data and categorical data separately: the numerical sample is generated by Gaussian diffusion models while the categorical sample is generated by Multinomial diffusion models. Since step-wise error comparison applies only to Gaussian models, we excluded categorical data when implementing the attack against TabDDPM.

The training process of both target models follows a similar setup of the TabSyn repository with minor modifications. We take 90\% of the dataset as the training data (member) for both models and keep the rest 10\% as the testing data (non-member). Before our experiments, we train the denoising model of TabDDPM on each training set with a batch size of 1024 for 50,000 epochs. For the TabSyn model, we first train the VAE on the training set for 4000 epochs with a batch size of 4096, then train the denoising model on the latent embeddings for 10,000 epochs with a batch size of 4096.

Since the \textbf{Default} dataset contains significantly more rows of records than the \textbf{Shoppers} dataset, we also include the result of the experiment where both target models are trained on the same number of rows to examine the effect of the size of the training data on the performance of our attack.

\subsection{Calculation of t-error} 
We strictly follow the equations listed in Section \ref{secmi} (Equation \ref{eq:terror}, \ref{eq:deterministic}, \ref{eq:f}) for calculating of the t-error. To obtain the noise \begin{math}\beta_t\end{math} at timestep t, we apply the cosine noise scheduler \cite{pmlr-v139-nichol21a} for the experiment with the TabDDPM model and the linear noise scheduler \cite{ho2020denoisingdiffusionprobabilisticmodels} for the experiment with the TabSyn model. The denoising model \begin{math}\epsilon_\theta\end{math} is loaded with the pre-trained parameter we obtained through the training process described in Section \ref{target}.

\subsection{Attack model} \label{attack}
The first attack model \begin{math}SecMI_{stat}\end{math} calculates a threshold such that test data with a lower score would be predicted as a member. In the original implementation of the SecMI, the score of an image data is calculated as the sum of the t-error at each pixel. Since we are working with tabular data, we changed the value of the score to the sum across all the columns for a single row of records.

The other attack model \begin{math}SecMI_{NNs}\end{math} involves the training of a neural network model. The model takes the column-wise t-error of each row of data as input along with the label (member or non-member). The job of this model is to predict a confidence score of membership for each row of input data. We keep the structure of the neural network model for image data unchanged besides that every 2D layer is changed to a 1D layer so that it fits tabular data (Table \ref{tab:block} and \ref{tab:NNs}).

\begin{table}
    \centering
    \begin{tabular}{|c|c|}
        \hline
        Layer & Parameter\\
         \hline\hline
        Conv1d & kernel size=3, stride=1, padding=1 \\ 
         \hline
        BatchNorm1d & - \\
        \hline
        Conv1d & kernel size=3, stride=1, padding=1 \\
        \hline
        BatchNorm1d & -\\
        \hline
    \end{tabular}
    \caption{Layer structure of a basic block in the neural network model.}
    \label{tab:block}
\end{table}

\begin{table}
    \centering
    \begin{tabular}{|c|c|}
        \hline
        Layer & Parameter\\
         \hline\hline
        Conv1d & kernel size=3, stride=1, padding=1 \\ 
         \hline
        BatchNorm1d & - \\
        \hline
        Basic block & - \\
        \hline
        Basic block & - \\
        \hline
        Basic block & - \\
        \hline
        Basic block & - \\
        \hline
        Linear & - \\
        \hline
    \end{tabular}
    \caption{Layer structure of the neural network model. The input shape and the output shape of this model depend on the size of the training tabular data.}
    \label{tab:NNs}
\end{table}

20\% of the data is used for training the neural network and 80\% for testing. The threshold for this approach is determined based on the membership likelihood output by the model. The sample being predicted with a higher score will be labeled as a member by our attack model.

\subsection{Evaluation}
We first adopt the Receiver Operating Characteristic (ROC) curve to measure the performance of the attack model,
which is commonly used for illustrating the performance of a binary
classifier model. The x-axis of the graph is the False Positive Rate (FPR) while the y-axis is the True Positive Rate (TPR), representing the trade-off between them. Here the False Positive is the case where our attack model classified a non-member as a member, and the True Positive is the case where the attack model classified a member as a member correctly. Each data point on the plotted line represented the FPR and the TPR of the attack model with a certain threshold. The score of an attack is calculated as the Area Under the Curve (AUC). A score of 1 means that the prediction is 100\% accurate, while 0.5 would be close to a random guess, meaning the attack failed.

However, \citeauthor{carlini2022membershipinferenceattacksprinciples} argued that such method with the use of a ROC curve only reflects the performance of the attack model on average cases, but in practice, we are more interested in the worst-case scenario. Moreover, a high FPR means that the prediction from the attack model is highly unreliable, so it is recommended that we instead look at the TPR at a very low FPR. As a result, in our experiment, we will be also looking at the TPR at 1\% FPR and at 0.1\% FPR.

\section{Experiment}
\subsection{Comparison of t-error} \label{terror_experiment}
Before performing the attack, we verify if the assumption of t-errors holds for synthetic tabular data. To do that, we first adopt the method from the SecMI paper where they sum up all pixel-wise t-errors and divide the value for non-member by the value for member. For tabular data, we sum up t-errors across every column. 

\begin{figure}[ht]
\centering
\begin{subfigure}{.5\textwidth}
  \centering
  \includegraphics[width=.9\linewidth]{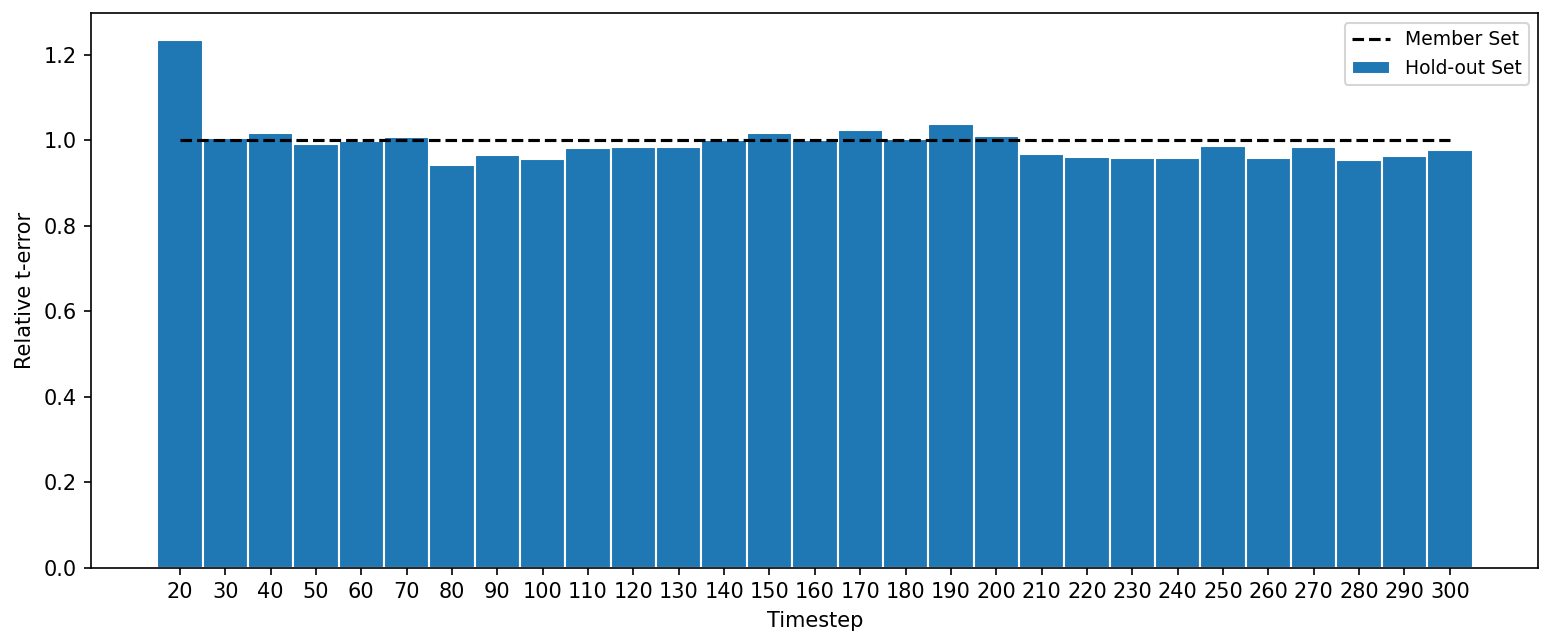}
  \caption{t-error comparison on Shoppers.}
\end{subfigure}
\begin{subfigure}{.5\textwidth}
  \centering
  \includegraphics[width=.9\linewidth]{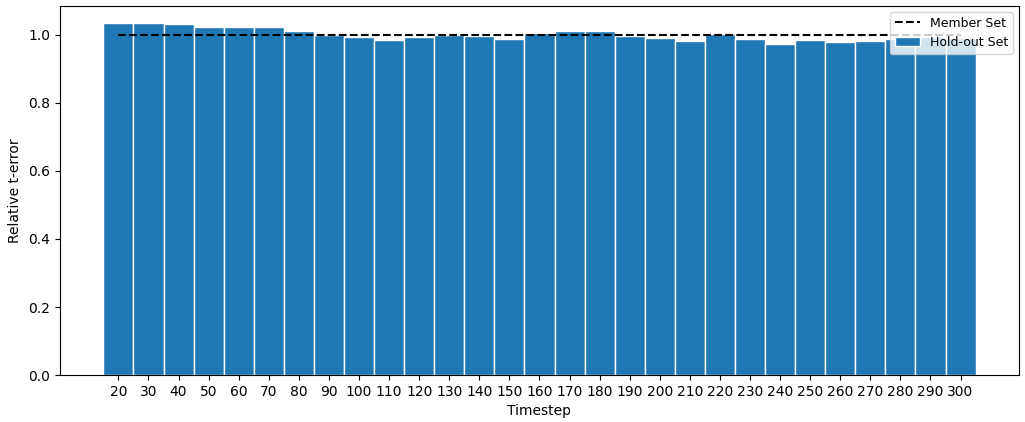}
  \caption{t-error comparison on Default.}
\end{subfigure}
\caption{\label{fig:t_sum_ddpm} TabDDPM t-error of non-member relative to member over different timestep from 20 to 300.}
\end{figure}

When the model is trained on the Default dataset, the t-error of non-members is slightly higher than the t-error of members from timestep 20 to 80 by less than 10\%, as shown in Figure \ref{fig:t_sum_ddpm}, which supports the assumption that the denoising model retains more "memory" at a smaller timestep so that it makes more accurate noise prediction when the input data is a member. It also shows that the t-error of non-members is approximately 1.2 times the t-error of members at timestep 20 for the Shoppers dataset, however, this difference becomes less significant from timestep 30. Since tabular data differs from pixel data, this result alone is insufficient to conclude that the assumption does not hold at timesteps greater than 20 for the Shoppers dataset. To further investigate, we compared the t-error across columns.

\begin{figure}[ht]
\centering
\begin{subfigure}{.5\textwidth}
  \centering
  \includegraphics[width=.9\linewidth]{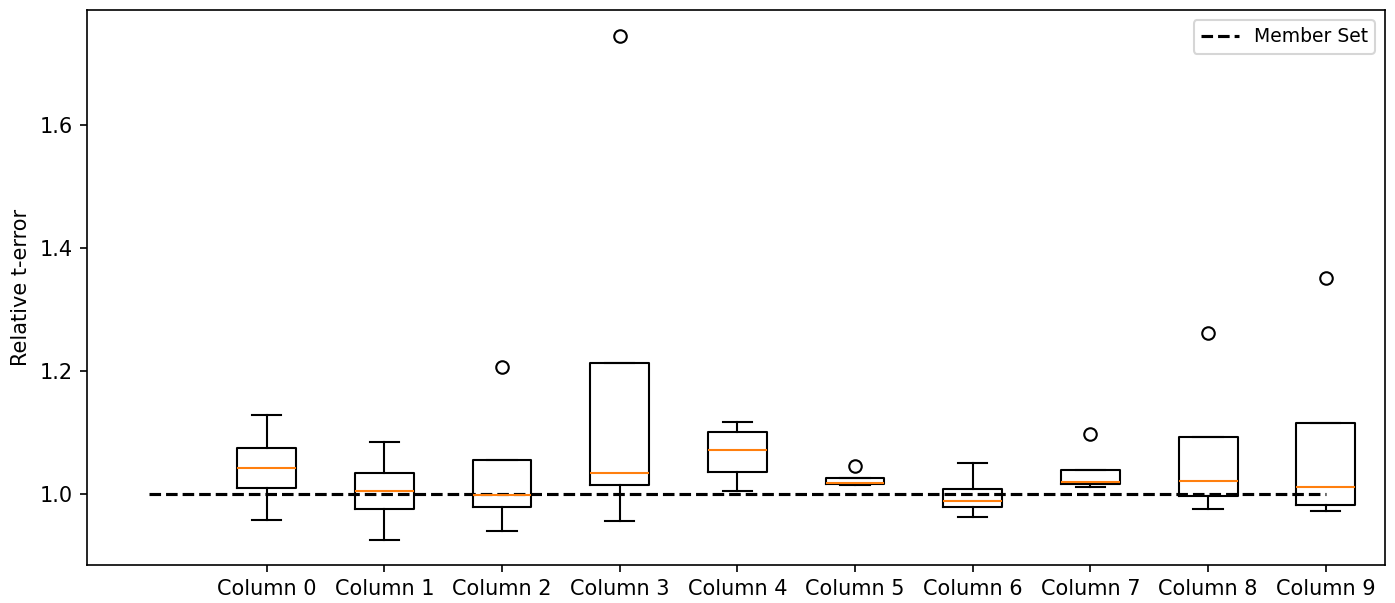}
  \caption{t-error comparison on Shoppers.}
\end{subfigure}
\begin{subfigure}{.5\textwidth}
  \centering
  \includegraphics[width=.9\linewidth]{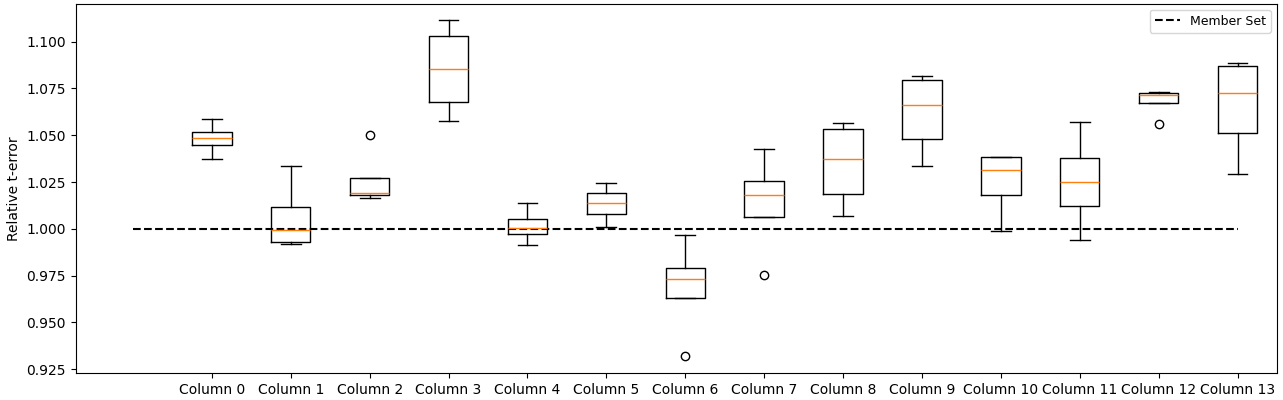}
  \caption{t-error comparison on Default.}
\end{subfigure}
\caption{\label{fig:column} Column-wise t-error of non-member relative to member at timestep from 20 to 50.}
\end{figure}

In Figure \ref{fig:column}, we observe that while the relative t-errors for some columns are close to 1, other columns, such as the column 4 of the Shoppers dataset and the column 3 of the Default dataset, show consistently higher t-errors for non-members. This potentially suggests that certain columns are more likely to overfit than others. 

\begin{figure}[ht]
\centering
\begin{subfigure}{.5\textwidth}
  \centering
  \includegraphics[width=.9\linewidth]{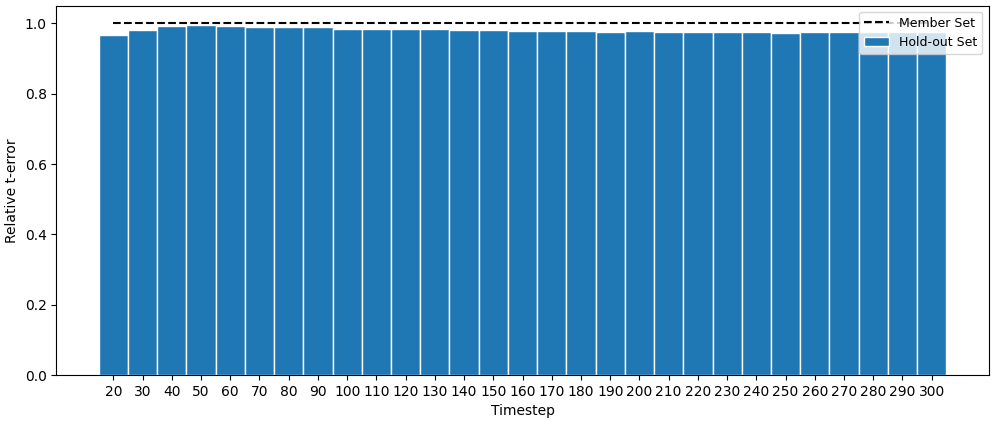}
  \caption{t-error comparison on Shoppers.}
\end{subfigure}
\begin{subfigure}{.5\textwidth}
  \centering
  \includegraphics[width=.9\linewidth]{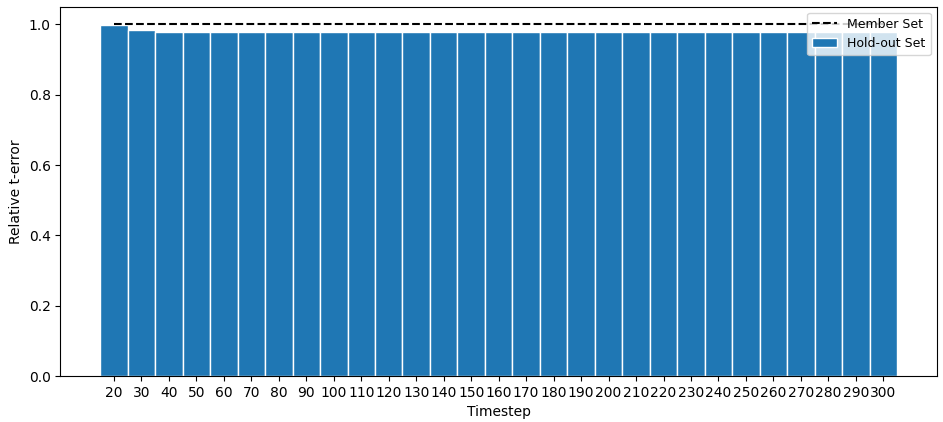}
  \caption{t-error comparison on Default.}
\end{subfigure}
\caption{\label{fig:t_sum_syn} TabSyn t-error of non-member relative to member over different timestep from 20 to 300.}
\end{figure}

The pattern observed in the t-error comparison of TabDDPM is notably different from that of TabSYN (Figure \ref{fig:t_sum_syn}), as the ratio of non-member to member remains almost the same across every timestep, with the t-error of non-member being constantly lower than that of member. This result is contrary to the assumption we mentioned earlier in Section \ref{secmi}, thus the SecMI which is based on this assumption would not be an appropriate attack model for TabSyn.

\begin{figure}[ht]
\centering
\begin{subfigure}{.25\textwidth}
  \centering
  \includegraphics[width=.9\linewidth]{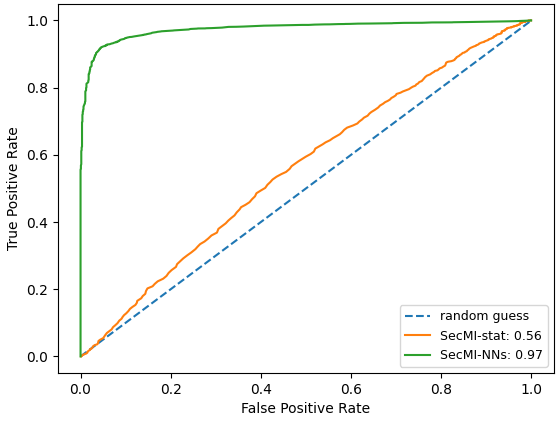}
  \caption{ROC on Shoppers.}
\end{subfigure}%
\begin{subfigure}{.25\textwidth}
  \centering
  \includegraphics[width=.9\linewidth]{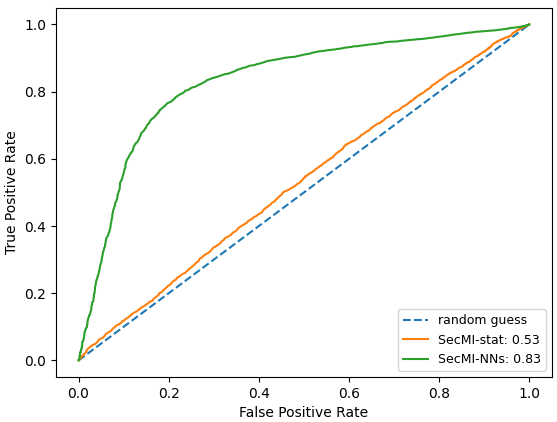}
  \caption{ROC on Default.}
\end{subfigure}
\begin{subfigure}{.25\textwidth}
  \centering
  \includegraphics[width=.9\linewidth]{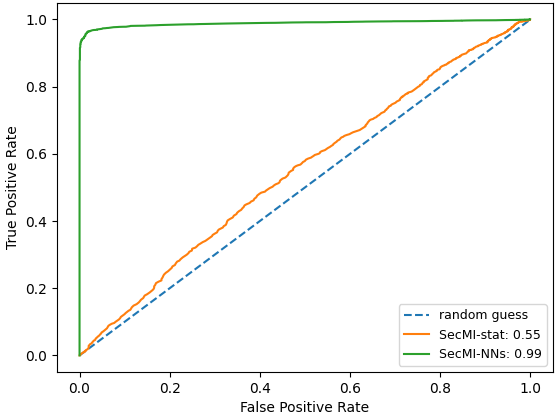}
  \caption{ROC on Default\begin{math}^{\prime}\end{math}}
\end{subfigure}
\caption{\label{fig:roc_tabddpm} ROC curves at timestep 50 for the attacks against TabDDPM.}
\end{figure}

\begin{table}
    \centering
    \begin{tabular}{|c|c|c|}
        \hline
         & TPR @ 1\% FPR & TPR @ 0.1\% FPR\\
        \hline\hline
        Shoppers & 75.4\% & 55.8\%\\
         \hline
        Default & 5.7\% & 0.1\%\\
         \hline
        Default\begin{math}^{\prime}\end{math} & 94.4\% & 88.0\%\\
         \hline
    \end{tabular}
    \caption{The TPR at low FPR of \begin{math}SecMI_{NNs}\end{math} on TabDDPM.}
    \label{tab:low_fpr_tabddpm}
\end{table}

\subsection{Results of attacks against TabDDPM}
In this section we apply the attack algorithm to the denoising models of TabDDPM.
Figure \ref{fig:roc_tabddpm} shows that \begin{math}SecMI_{NNs}\end{math} achieved an AUC of 97\% on TabDDPM trained on Shoppers and 83\% on Default, however the scores of both \begin{math}SecMI_{stat}\end{math} attacks are close to a random guess. Since \begin{math}SecMI_{stat}\end{math} takes the sum of t-errors as input whereas \begin{math}SecMI_{NNs}\end{math} learns from column-wise t-errors, this large gap in performance could be explained by the observation that, although the overall difference in t-errors at timestep 50 is less than 10\%, certain columns amplify this difference. 

If we look at the worst case scenario quantified with the TPR at a low FPR displayed in Table \ref{tab:low_fpr_tabddpm}, the attack against TabDDPM trained on Shoppers proves to be successful with a TPR of a 75.4\% at 1\% FPR and a TPR of a 55.8\% at 0.1\% FPR. This result can be interpreted in this way: There exist a threshold such that our attack could successfully leak the presence of 75.4\% of the data in the training set with 1\% of the data not in the training set being mistakenly identified as part of the training dataset. However, there is a dramatic drop in performance when we switch to the Default dataset, with only 5.7\% TPR at 1\% FPR and 0.1\% TPR at 0.1\% FPR. This shows that even if the AUC score on Default is high, in practice the attacker could only confidently identify a small portion from the entire dataset as the member. 

Since two datasets are different in size, we also perform the attack using Default\begin{math}^{\prime}\end{math} which is the Default dataset reduced to the same size as the Shoppers dataset. Note that both Shoppers and Default\begin{math}^{\prime}\end{math} contains 12,330 rows of data while Default contains 30,000 rows in total, and 90\% of the data from Shoppers, Default and Default\begin{math}^{\prime}\end{math} are used as the training data for the TabDDPM model. As a result of a reduction in the training size, the AUC score of the \begin{math}SecMI_{NNs}\end{math} is shockingly increased to 99\%, with the TPR at 1\% FPR increased to 94.4\% and 88.0\% at 0.1\% FPR. Although previous results have shown that our attack is more effective when the target model is trained on Shoppers, if we strictly resize two datasets to be exactly the same amount, the attack would actually achieve a better performance on Default.

\begin{figure}[ht]
\centering
\begin{subfigure}{.25\textwidth}
  \centering
  \includegraphics[width=.9\linewidth]{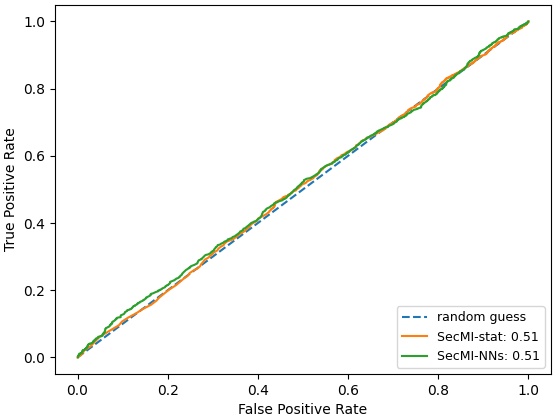}
  \caption{ROC on Shoppers.}
\end{subfigure}%
\begin{subfigure}{.25\textwidth}
  \centering
  \includegraphics[width=.9\linewidth]{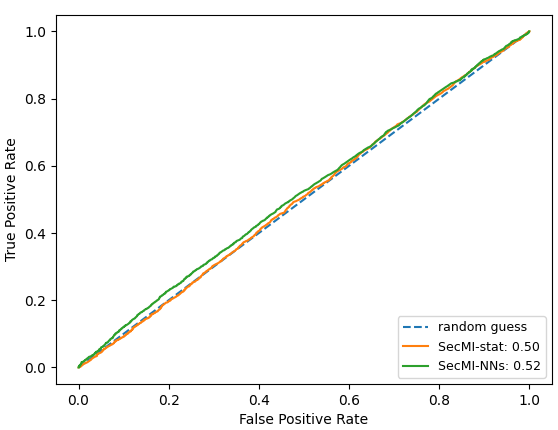}
  \caption{ROC on Default.}
\end{subfigure}
\begin{subfigure}{.25\textwidth}
  \centering
  \includegraphics[width=.9\linewidth]{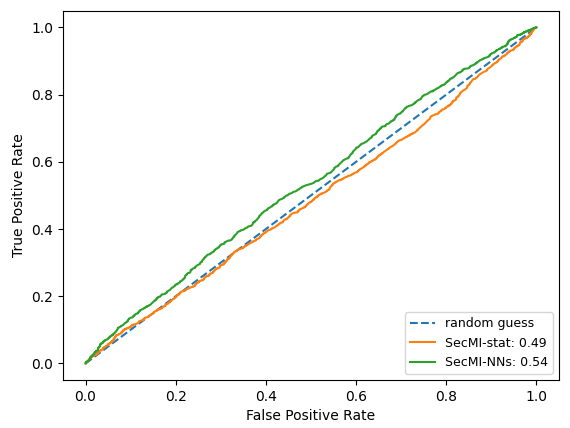}
  \caption{ROC on Default\begin{math}^{\prime}\end{math}}
\end{subfigure}
\caption{\label{fig:roc_tabsyn} ROC curves at timestep 50 for the attacks against TabSyn.}
\end{figure}

\begin{table}
    \centering
    \begin{tabular}{|c|c|c|}
        \hline
         & TPR @ 1\% FPR & TPR @ 0.1\% FPR\\
        \hline\hline
        Shoppers & 1.4\% & 0.0\%\\
         \hline
        Default & 1.7\% & 0.2\%\\
         \hline
        Default\begin{math}^{\prime}\end{math} & 1.3\% & 0.2\%\\
         \hline
    \end{tabular}
    \caption{The TPR at low FPR of \begin{math}SecMI_{NNs}\end{math} against TabSyn.}
    \label{tab:low_fpr_tabsyn}
\end{table}

\subsection{Results of attacks against TabSyn}
In this section we apply the attack algorithm to the denoising models of TabSyn. The behavior of both \begin{math}SecMI_{stat}\end{math} and \begin{math}SecMI_{NNs}\end{math} is very similar to a random guess, as shown in Figure \ref{fig:roc_tabsyn} with the score of AUC lower than 51\%. At a low FPR, the value of TPR is close to the FPR with a difference less than 1\%, implying that the result of our attack against TabSyn is not reliable. Although a reduction in the training size offered the attacker a great advantage in attacking TabDDPM, for the attacks against TabSyn on the Default\begin{math}^{\prime}\end{math} dataset, no significant difference is observed in the performance comparing to the attacks on the Default dataset.

\section{Discussion}
\subsection{Privacy evaluation methods}
\citeauthor{zhang2024mixedtypetabulardatasynthesis} adopts DCR as the evaluation metrics to assess the privacy risks of TabDDPM and TabSyn. Here, DCR stands for the distance between a synthetic sample and the closest real sample to that synthetic sample. They first calculate the DCR between the synthetic dataset and the training(member)/non-member set, then compare the distribution of the DCR values. As stated by them, the synthetic dataset is considered almost identical to the training set if the DCR for the training set is close to zero. On the other hand, a similar distribution of DCR between synthetic and training sample and between synthetic and non-member sample indicates low privacy risks. To quantify the privacy risks, they calculate DCR score which represent the probability that a synthetic sample is closer to the training set than the non-member set, where a score closer to 50\% implies that the synthetic dataset is equally close to both sets. The DCR score obtained on Default and Shoppers is reported in Table \ref{tab:dcr}. The values are closed to 50\%, except for the DCR score for TabDDPM on Shoppers which hits above 60\%. With this table, \citeauthor{zhang2024mixedtypetabulardatasynthesis} claimed that both TabDDPM and TabSyn do not suffer from privacy issues. 

However, is DCR really an effective methods for evaluating the privacy risks of diffusion model? The intuition of adopting this method is that, if a synthetic sample is too close to a training sample to the point that the attacker can recognize that the synthetic sample is related to that training sample. For example, when we are adding noises to the real data as a form of privacy protection, each record from the 'fake' dataset corresponds to one from the real dataset, and if they are too similar to each other the attacker could potentially infer the original data just by looking at the 'fake' dataset. However, this is not the case for the synthetic data generated by diffusion models. Diffusion models learn how to denoise the data at each timestep through the training process and then generate data samples from pure noise through the denoising process, thus the generated samples only maintain similar statistical properties and are not directly related to the training samples. In fact, the denoising model is the one which is highly influenced by the training data, and as a result, it is the main target of our attacks. Although our results does seem to align with the DCR scores, as the TabDDPM on Shoppers has notably higher DCR scores than the others and it also has the highest TPR at a low FPR, instead of reporting the similarity between the synthetic data and the real data, our attacks could tell you what percentage of the member data are in danger of being inferred.

\begin{table}
    \centering
    \begin{tabular}{|c|c|c|}
        \hline
         & Shoppers & Default\\
         \hline\hline
        TabDDPM & 63.23\%±0.25 & 52.15\%±0.20\\
        \hline
        TabSyn & 52.90\%±0.22 & 51.20\%±0.18\\
        \hline
    \end{tabular}
    \caption{DCR scores reported by TabSyn. A score close to 50\% is considered better in terms of privacy protection.}
    \label{tab:dcr}
\end{table}

\subsection{Vulnerability factors}
In this section we investigate why is our attack only effective for certain datasets and diffusion models. A notable factor we observed from the experiments is the training size. It is known that MIAs benefit from overfitting \cite{yeom2018privacyriskmachinelearning}, and overfitting may happen when the size of training data is so small that it does not accurately represent the majority of all possible data in this domain. As shown in Table \ref{tab:low_fpr_tabddpm}, our attacks against TabDDPM on Default has a much lower TPR at a low FPR than on Shoppers, however when trained on the same dataset with a smaller size, the same attack method achieved a higher TPR at a low FPR than on Shoppers. Since no other factors have changed, we can conclude that using a smaller training set does result in greater privacy risks, making the diffusion model more vulnerable to MIAs.

The other factor is how the data is handled during the training process. TabDDPM normalizes every numerical column in the training sample and uses one-hot encoding to convert categorical columns into numerical columns when preparing the data for training the denoising model. Therefore, the denoising model of TabDDPM is still trained on the tabular data but with some values being modified. TabSyn handled the data differently, as they instead designed a VAE that encodes the tabular data as a probability distribution over the latent space. Then the whole training process of the denoising model takes place in the latent space instead of in the data space. Here the latent embedding of a record is more of a representation of relationships among all columns rather than each individual column. In Section \ref{terror_experiment}, Figure \ref{fig:t_sum_syn} suggests no significant overfitting for TabSyn as the t-error of member data is not higher than that of non-member. This observation might be due to the lack of significant differences between the latent embeddings of member and non-member samples; however, further investigation is needed to validate this.

\subsection{Limitations and future works}
Our study suffers from several limitations. First, due to the lack of computational resources, we investigate the performance of our MIAs on only two datasets, which do not represent the majority of tabular data. Our findings require confirmation from additional experiments on other datasets. 

Although SecMI fails to infer membership from TabSyn, this does not mean TabSyn does not suffer from privacy risks. Designing a MIA model targeting input data in the form of latent embeddings would be an interesting direction so that we could have an effective privacy evaluation method for latent diffusion models as well.

In practice, implementing SecMI is challenging. It requires access to the denoising model, meaning that if the model publisher only provides black-box access (where users can only input training data and receive synthetic output), our attack cannot be implemented. Therefore, SecMI should be viewed more as a privacy evaluation method rather than a real threat.

\section{Conclusion}
This study stems from the growing use of diffusion-based generative models for creating synthetic tabular data and the need to assess their vulnerability to privacy risks like MIAs. We explored the vulnerabilities of two recent models, TabDDPM and TabSyn, by implementing query-based MIAs and analyzing their effectiveness under varying conditions.

Our findings show that TabDDPM is more vulnerable to our attacks than TabSyn, especially with smaller training datasets. These results highlight the significant impact of model architecture on privacy resilience. Notably, our experiments emphasize the relationship between dataset size, overfitting, and privacy vulnerability.

Overall, our work shows the importance of assessing the privacy protection of diffusion models using appropriate metrics. We provide an example of the attack against diffusion-based synthetic tabular data as a form of privacy evaluation, and we hope our work will inspire the development of stronger attacks in the future. Furthermore, we hope our work will raise awareness of the need for privacy considerations in diffusion models.
\bibliographystyle{ACM-Reference-Format}
\bibliography{ref}


\begin{thebibliography}{20}


\ifx \showCODEN    \undefined \def \showCODEN     #1{\unskip}     \fi
\ifx \showDOI      \undefined \def \showDOI       #1{#1}\fi
\ifx \showISBNx    \undefined \def \showISBNx     #1{\unskip}     \fi
\ifx \showISBNxiii \undefined \def \showISBNxiii  #1{\unskip}     \fi
\ifx \showISSN     \undefined \def \showISSN      #1{\unskip}     \fi
\ifx \showLCCN     \undefined \def \showLCCN      #1{\unskip}     \fi
\ifx \shownote     \undefined \def \shownote      #1{#1}          \fi
\ifx \showarticletitle \undefined \def \showarticletitle #1{#1}   \fi
\ifx \showURL      \undefined \def \showURL       {\relax}        \fi
\providecommand\bibfield[2]{#2}
\providecommand\bibinfo[2]{#2}
\providecommand\natexlab[1]{#1}
\providecommand\showeprint[2][]{arXiv:#2}

\bibitem[Carlini et~al\mbox{.}(2022)]%
        {carlini2022membershipinferenceattacksprinciples}
\bibfield{author}{\bibinfo{person}{Nicholas Carlini}, \bibinfo{person}{Steve Chien}, \bibinfo{person}{Milad Nasr}, \bibinfo{person}{Shuang Song}, \bibinfo{person}{Andreas Terzis}, {and} \bibinfo{person}{Florian Tramer}.} \bibinfo{year}{2022}\natexlab{}.
\newblock \bibinfo{title}{Membership Inference Attacks From First Principles}.
\newblock
\newblock
\showeprint[arxiv]{2112.03570}~[cs.CR]
\urldef\tempurl%
\url{https://arxiv.org/abs/2112.03570}
\showURL{%
\tempurl}


\bibitem[Duan et~al\mbox{.}(2023)]%
        {duan2023diffusionmodelsvulnerablemembership}
\bibfield{author}{\bibinfo{person}{Jinhao Duan}, \bibinfo{person}{Fei Kong}, \bibinfo{person}{Shiqi Wang}, \bibinfo{person}{Xiaoshuang Shi}, {and} \bibinfo{person}{Kaidi Xu}.} \bibinfo{year}{2023}\natexlab{}.
\newblock \bibinfo{title}{Are Diffusion Models Vulnerable to Membership Inference Attacks?}
\newblock
\newblock
\showeprint[arxiv]{2302.01316}~[cs.CV]
\urldef\tempurl%
\url{https://arxiv.org/abs/2302.01316}
\showURL{%
\tempurl}


\bibitem[Emam et~al\mbox{.}(2011)]%
        {ElEmam2011}
\bibfield{author}{\bibinfo{person}{Khaled~El Emam}, \bibinfo{person}{David Buckeridge}, \bibinfo{person}{Robyn Tamblyn}, \bibinfo{person}{Bartha~Maria Knoppers}, \bibinfo{person}{Valerie Fineberg}, \bibinfo{person}{Susan Reid}, \bibinfo{person}{Farouk Jonker}, {and} \bibinfo{person}{Philip~M. Gold}.} \bibinfo{year}{2011}\natexlab{}.
\newblock \showarticletitle{The Re-identification Risk of Canadians from Longitudinal Demographics}.
\newblock \bibinfo{journal}{\emph{BMC Medical Informatics and Decision Making}}  \bibinfo{volume}{11} (\bibinfo{year}{2011}), \bibinfo{pages}{46}.
\newblock
\urldef\tempurl%
\url{https://doi.org/10.1186/1472-6947-11-46}
\showDOI{\tempurl}


\bibitem[Hayes et~al\mbox{.}(2018)]%
        {hayes2018loganmembershipinferenceattacks}
\bibfield{author}{\bibinfo{person}{Jamie Hayes}, \bibinfo{person}{Luca Melis}, \bibinfo{person}{George Danezis}, {and} \bibinfo{person}{Emiliano~De Cristofaro}.} \bibinfo{year}{2018}\natexlab{}.
\newblock \bibinfo{title}{LOGAN: Membership Inference Attacks Against Generative Models}.
\newblock
\newblock
\showeprint[arxiv]{1705.07663}~[cs.CR]
\urldef\tempurl%
\url{https://arxiv.org/abs/1705.07663}
\showURL{%
\tempurl}


\bibitem[Ho et~al\mbox{.}(2020)]%
        {ho2020denoisingdiffusionprobabilisticmodels}
\bibfield{author}{\bibinfo{person}{Jonathan Ho}, \bibinfo{person}{Ajay Jain}, {and} \bibinfo{person}{Pieter Abbeel}.} \bibinfo{year}{2020}\natexlab{}.
\newblock \bibinfo{title}{Denoising Diffusion Probabilistic Models}.
\newblock
\newblock
\showeprint[arxiv]{2006.11239}~[cs.LG]
\urldef\tempurl%
\url{https://arxiv.org/abs/2006.11239}
\showURL{%
\tempurl}


\bibitem[Holmes et~al\mbox{.}(2019)]%
        {holmes2019artificial}
\bibfield{author}{\bibinfo{person}{Wayne Holmes}, \bibinfo{person}{Maya Bialik}, {and} \bibinfo{person}{Charles Fadel}.} \bibinfo{year}{2019}\natexlab{}.
\newblock \bibinfo{booktitle}{\emph{Artificial intelligence in education promises and implications for teaching and learning}}.
\newblock \bibinfo{publisher}{Center for Curriculum Redesign}.
\newblock


\bibitem[Jordon et~al\mbox{.}(2019)]%
        {Jordon2019}
\bibfield{author}{\bibinfo{person}{James Jordon}, \bibinfo{person}{Jinsung Yoon}, {and} \bibinfo{person}{Mihaela van~der Schaar}.} \bibinfo{year}{2019}\natexlab{}.
\newblock \showarticletitle{{PATE-GAN}: Generating Synthetic Data with Differential Privacy Guarantees}. In \bibinfo{booktitle}{\emph{International Conference on Learning Representations}}.
\newblock
\urldef\tempurl%
\url{https://arxiv.org/abs/1806.06668}
\showURL{%
\tempurl}


\bibitem[Kotelnikov et~al\mbox{.}(2022)]%
        {kotelnikov2022tabddpmmodellingtabulardata}
\bibfield{author}{\bibinfo{person}{Akim Kotelnikov}, \bibinfo{person}{Dmitry Baranchuk}, \bibinfo{person}{Ivan Rubachev}, {and} \bibinfo{person}{Artem Babenko}.} \bibinfo{year}{2022}\natexlab{}.
\newblock \bibinfo{title}{TabDDPM: Modelling Tabular Data with Diffusion Models}.
\newblock
\newblock
\showeprint[arxiv]{2209.15421}~[cs.LG]
\urldef\tempurl%
\url{https://arxiv.org/abs/2209.15421}
\showURL{%
\tempurl}


\bibitem[Murakonda and Shokri(2020)]%
        {murakonda2020mlprivacymeteraiding}
\bibfield{author}{\bibinfo{person}{Sasi~Kumar Murakonda} {and} \bibinfo{person}{Reza Shokri}.} \bibinfo{year}{2020}\natexlab{}.
\newblock \bibinfo{title}{ML Privacy Meter: Aiding Regulatory Compliance by Quantifying the Privacy Risks of Machine Learning}.
\newblock
\newblock
\showeprint[arxiv]{2007.09339}~[cs.CR]
\urldef\tempurl%
\url{https://arxiv.org/abs/2007.09339}
\showURL{%
\tempurl}


\bibitem[Narayanan and Shmatikov(2008)]%
        {Narayanan2008}
\bibfield{author}{\bibinfo{person}{Arvind Narayanan} {and} \bibinfo{person}{Vitaly Shmatikov}.} \bibinfo{year}{2008}\natexlab{}.
\newblock \showarticletitle{Robust De-anonymization of Large Sparse Datasets}. In \bibinfo{booktitle}{\emph{Proceedings of the 2008 IEEE Symposium on Security and Privacy}}. \bibinfo{publisher}{IEEE Computer Society}, \bibinfo{address}{USA}, \bibinfo{pages}{111--125}.
\newblock
\showISBNx{9780769531687}
\urldef\tempurl%
\url{https://doi.org/10.1109/SP.2008.33}
\showDOI{\tempurl}


\bibitem[Nichol and Dhariwal(2021)]%
        {pmlr-v139-nichol21a}
\bibfield{author}{\bibinfo{person}{Alexander~Quinn Nichol} {and} \bibinfo{person}{Prafulla Dhariwal}.} \bibinfo{year}{2021}\natexlab{}.
\newblock \showarticletitle{Improved Denoising Diffusion Probabilistic Models}. In \bibinfo{booktitle}{\emph{Proceedings of the 38th International Conference on Machine Learning}} \emph{(\bibinfo{series}{Proceedings of Machine Learning Research}, Vol.~\bibinfo{volume}{139})}, \bibfield{editor}{\bibinfo{person}{Marina Meila} {and} \bibinfo{person}{Tong Zhang}} (Eds.). \bibinfo{publisher}{PMLR}, \bibinfo{pages}{8162--8171}.
\newblock
\urldef\tempurl%
\url{https://proceedings.mlr.press/v139/nichol21a.html}
\showURL{%
\tempurl}


\bibitem[Pang et~al\mbox{.}(2024)]%
        {pang2024clavaddpmmultirelationaldatasynthesis}
\bibfield{author}{\bibinfo{person}{Wei Pang}, \bibinfo{person}{Masoumeh Shafieinejad}, \bibinfo{person}{Lucy Liu}, {and} \bibinfo{person}{Xi He}.} \bibinfo{year}{2024}\natexlab{}.
\newblock \bibinfo{title}{ClavaDDPM: Multi-relational Data Synthesis with Cluster-guided Diffusion Models}.
\newblock
\newblock
\showeprint[arxiv]{2405.17724}~[cs.AI]
\urldef\tempurl%
\url{https://arxiv.org/abs/2405.17724}
\showURL{%
\tempurl}


\bibitem[Park et~al\mbox{.}(2018)]%
        {Park_2018}
\bibfield{author}{\bibinfo{person}{Noseong Park}, \bibinfo{person}{Mahmoud Mohammadi}, \bibinfo{person}{Kshitij Gorde}, \bibinfo{person}{Sushil Jajodia}, \bibinfo{person}{Hongkyu Park}, {and} \bibinfo{person}{Youngmin Kim}.} \bibinfo{year}{2018}\natexlab{}.
\newblock \showarticletitle{Data synthesis based on generative adversarial networks}.
\newblock \bibinfo{journal}{\emph{Proceedings of the VLDB Endowment}} \bibinfo{volume}{11}, \bibinfo{number}{10} (\bibinfo{date}{June} \bibinfo{year}{2018}), \bibinfo{pages}{1071–1083}.
\newblock
\showISSN{2150-8097}
\urldef\tempurl%
\url{https://doi.org/10.14778/3231751.3231757}
\showDOI{\tempurl}


\bibitem[Rajkomar et~al\mbox{.}(2019)]%
        {rajkomar2019machine}
\bibfield{author}{\bibinfo{person}{Alvin Rajkomar}, \bibinfo{person}{Jeffrey Dean}, {and} \bibinfo{person}{Isaac Kohane}.} \bibinfo{year}{2019}\natexlab{}.
\newblock \showarticletitle{Machine learning in medicine}.
\newblock \bibinfo{journal}{\emph{New England Journal of Medicine}} \bibinfo{volume}{380}, \bibinfo{number}{14} (\bibinfo{year}{2019}), \bibinfo{pages}{1347--1358}.
\newblock


\bibitem[Shokri et~al\mbox{.}(2017)]%
        {shokri2017membershipinferenceattacksmachine}
\bibfield{author}{\bibinfo{person}{Reza Shokri}, \bibinfo{person}{Marco Stronati}, \bibinfo{person}{Congzheng Song}, {and} \bibinfo{person}{Vitaly Shmatikov}.} \bibinfo{year}{2017}\natexlab{}.
\newblock \bibinfo{title}{Membership Inference Attacks against Machine Learning Models}.
\newblock
\newblock
\showeprint[arxiv]{1610.05820}~[cs.CR]
\urldef\tempurl%
\url{https://arxiv.org/abs/1610.05820}
\showURL{%
\tempurl}


\bibitem[Sweeney(1997)]%
        {Sweeney1997}
\bibfield{author}{\bibinfo{person}{Latanya Sweeney}.} \bibinfo{year}{1997}\natexlab{}.
\newblock \showarticletitle{Weaving Technology and Policy Together to Maintain Confidentiality}.
\newblock \bibinfo{journal}{\emph{Journal of Law, Medicine \& Ethics}} \bibinfo{volume}{25}, \bibinfo{number}{2-3} (\bibinfo{year}{1997}), \bibinfo{pages}{98--110}.
\newblock
\urldef\tempurl%
\url{https://doi.org/10.1111/j.1748-720x.1997.tb01885.x}
\showDOI{\tempurl}


\bibitem[Yeom et~al\mbox{.}(2018)]%
        {yeom2018privacyriskmachinelearning}
\bibfield{author}{\bibinfo{person}{Samuel Yeom}, \bibinfo{person}{Irene Giacomelli}, \bibinfo{person}{Matt Fredrikson}, {and} \bibinfo{person}{Somesh Jha}.} \bibinfo{year}{2018}\natexlab{}.
\newblock \bibinfo{title}{Privacy Risk in Machine Learning: Analyzing the Connection to Overfitting}.
\newblock
\newblock
\showeprint[arxiv]{1709.01604}~[cs.CR]
\urldef\tempurl%
\url{https://arxiv.org/abs/1709.01604}
\showURL{%
\tempurl}


\bibitem[Zhang et~al\mbox{.}(2024)]%
        {zhang2024mixedtypetabulardatasynthesis}
\bibfield{author}{\bibinfo{person}{Hengrui Zhang}, \bibinfo{person}{Jiani Zhang}, \bibinfo{person}{Balasubramaniam Srinivasan}, \bibinfo{person}{Zhengyuan Shen}, \bibinfo{person}{Xiao Qin}, \bibinfo{person}{Christos Faloutsos}, \bibinfo{person}{Huzefa Rangwala}, {and} \bibinfo{person}{George Karypis}.} \bibinfo{year}{2024}\natexlab{}.
\newblock \bibinfo{title}{Mixed-Type Tabular Data Synthesis with Score-based Diffusion in Latent Space}.
\newblock
\newblock
\showeprint[arxiv]{2310.09656}~[cs.LG]
\urldef\tempurl%
\url{https://arxiv.org/abs/2310.09656}
\showURL{%
\tempurl}


\bibitem[Zhao et~al\mbox{.}(2022)]%
        {zhao2022ctabganenhancingtabulardata}
\bibfield{author}{\bibinfo{person}{Zilong Zhao}, \bibinfo{person}{Aditya Kunar}, \bibinfo{person}{Robert Birke}, {and} \bibinfo{person}{Lydia~Y. Chen}.} \bibinfo{year}{2022}\natexlab{}.
\newblock \bibinfo{title}{CTAB-GAN+: Enhancing Tabular Data Synthesis}.
\newblock
\newblock
\showeprint[arxiv]{2204.00401}~[cs.LG]
\urldef\tempurl%
\url{https://arxiv.org/abs/2204.00401}
\showURL{%
\tempurl}


\bibitem[Zhao et~al\mbox{.}(2021)]%
        {zhao2021ctabganeffectivetabledata}
\bibfield{author}{\bibinfo{person}{Zilong Zhao}, \bibinfo{person}{Aditya Kunar}, \bibinfo{person}{Hiek~Van der Scheer}, \bibinfo{person}{Robert Birke}, {and} \bibinfo{person}{Lydia~Y. Chen}.} \bibinfo{year}{2021}\natexlab{}.
\newblock \bibinfo{title}{CTAB-GAN: Effective Table Data Synthesizing}.
\newblock
\newblock
\showeprint[arxiv]{2102.08369}~[cs.LG]
\urldef\tempurl%
\url{https://arxiv.org/abs/2102.08369}
\showURL{%
\tempurl}


\end{thebibliography}

\end{document}